\documentclass{INTERSPEECH2023}
\usepackage{amsmath,graphicx,amsfonts,amssymb,stfloats,booktabs,multirow,bbding,pifont,url,color,xcolor,makecell,cite}
\usepackage{hyperref}

\newcommand{\cmark}{\ding{51}}%
\newcommand{\xmark}{\ding{55}}%

\interspeechcameraready

\title{Dual-Path Style Learning for End-to-End Noise-Robust Speech Recognition}

\name{Yuchen Hu, Nana Hou$^\ast$\thanks{$^\ast$Nana Hou contributed to this work before leaving Nanyang Technological University, Singapore.}, Chen Chen, Eng Siong Chng}

\address{School of Computer Science and Engineering, Nanyang Technological University, Singapore}
\email{yuchen005@e.ntu.edu.sg}

\begin{document}

\maketitle

\begin{abstract}
Automatic speech recognition (ASR) systems degrade significantly under noisy conditions.
Recently, speech enhancement (SE) is introduced as front-end to reduce noise for ASR, but it also suppresses some important speech information, \textit{i.e.}, over-suppression. 
To alleviate this, we propose a dual-path style learning approach for end-to-end noise-robust speech recognition (DPSL-ASR). 
Specifically, we first introduce clean speech feature along with the fused feature from IFF-Net as dual-path inputs to recover the suppressed information. 
Then, we propose style learning to map the fused feature close to clean feature, in order to learn latent speech information from the latter, \textit{i.e.}, clean ``speech style''. 
Furthermore, we also minimize the distance of final ASR outputs in two paths to improve noise-robustness.
Experiments show that the proposed approach achieves relative word error rate (WER) reductions of 10.6\% and 8.6\% over the best IFF-Net baseline, on RATS and CHiME-4 datasets respectively\footnote{https://github.com/YUCHEN005/DPSL-ASR}.
\end{abstract}


\noindent\textbf{Index Terms}: Dual-path style learning, consistency loss, noise-robust speech recognition, over-suppression problem

\section{Introduction}
\label{sec:intro}
Automatic speech recognition (ASR) has achieved a great success with recent advances of deep learning techniques~\cite{graves2006connectionist, graves2012sequence, chan2016listen, vaswani2017attention, watanabe2017hybrid,dong2018speech,radford2022robust}, which has been widely used in practical applications. 
However, it is still a challenging task when put under extremely noisy conditions, particularly in radio communication speech~\cite{hou2020multi}, which are distorted by ambient noise as well as communication channel due to limited transfer bandwidth.

Prior works usually introduce speech enhancement (SE)~\cite{Wang2014on,pascual2017segan,subramanian2019speech,liu2019jointly,wang2020complex,zheng2021interactive,pandey2021dual} as a front-end pre-processing module to reduce the additive noise and improve speech quality for downstream ASR task. 
However, recent works~\cite{mporas2010speech, loizou2011reasons, wang2020bridging} observe that the SE processing could not always improve ASR performance, as some important speech content information for ASR in original noisy speech~\cite{hu2022interactive} are also reduced by SE, together with the additive noise. Such over-suppression problem is usually undetected during the speech enhancement stage, but could significantly degrade the performance of downstream ASR task.

To alleviate this issue, prior work~\cite{hu2022interactive} proposes an interactive feature fusion network (IFF-Net) with joint SE-ASR framework. Specifically, it designs a CNN-Attention based architecture to interactively fuse the over-suppressed enhanced speech with original noisy speech, in order to recover the lost information in enhanced speech for downstream ASR. It achieves improvements but over-suppression problem can still be observed. 

In this paper, we propose a novel dual-path style learning approach for end-to-end noise-robust automatic speech recognition (DPSL-ASR) to further alleviate over-suppression problem. 
Specifically, we first introduce clean speech feature along with the fused feature from IFF-Net~\cite{hu2022interactive} as dual-path inputs to the back-end ASR module, where the clean feature can complement the suppressed information in corresponding fused feature. 
We then propose a style learning method to map fused feature close to clean feature, in order to learn latent speech information from the latter, \textit{i.e.}, clean ``speech style''.
Furthermore, we employ consistency loss to minimize the distance of ASR outputs in two paths to improve noise-robustness.
Experimental results show that the proposed approach significantly outperforms the best IFF-Net baseline, and further visualizations of intermediate embeddings indicate that DPSL-ASR can effectively recover the over-suppressed information by SE. 


\section{DPSL-ASR Architecture}
\label{sec:dpsl_asr}

\subsection{Overview}
\label{ssec:overview}

\begin{figure*}[ht]
  \centering
  \includegraphics[width=0.8\textwidth]{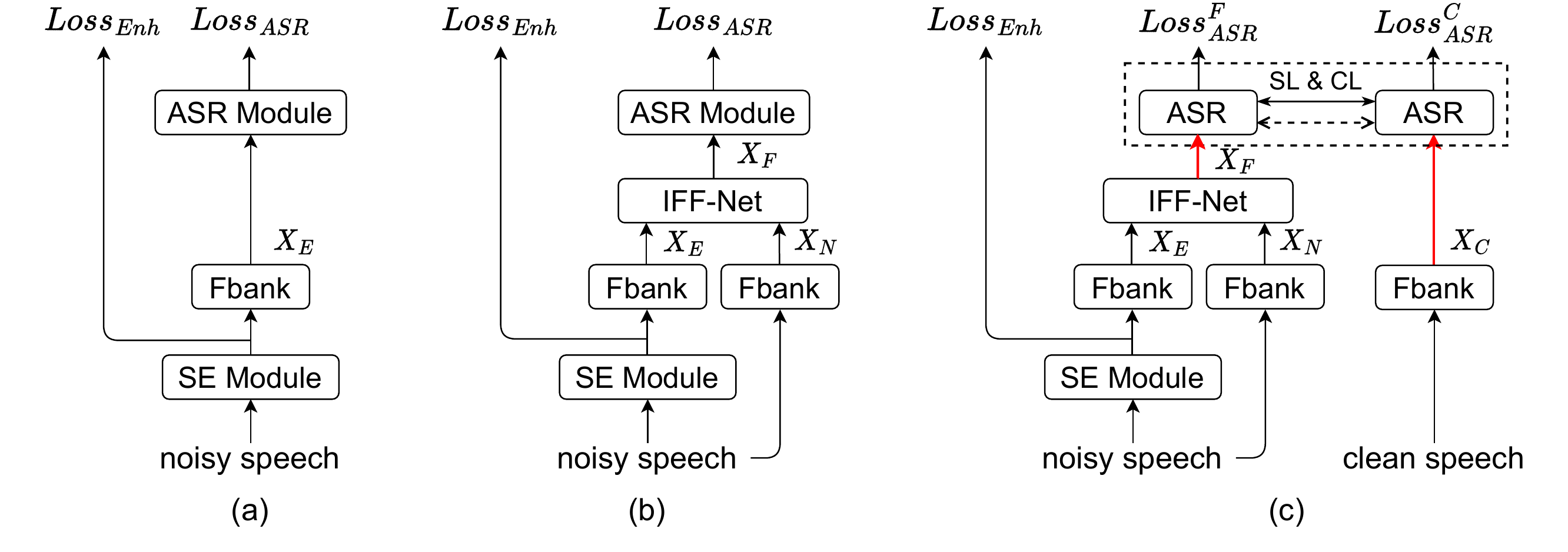}
  \vspace{-0.1cm}
  \caption{Block diagrams of (a) joint SE-ASR approach, (b) IFF-Net baseline, (c) the proposed DPSL-ASR approach. 
  The red arrows in (c) highlight the dual-path inputs.
  ``SL \& CL'' in the black dashed box denotes the style learning and consistency loss.
  The black dashed arrow between ASR modules denotes sharing parameters, so that DPSL-ASR contains same number of parameters as IFF-Net.
  }\label{fig1}
  \vspace{-0.25cm}
\end{figure*}

In this work, we propose a novel dual-path style learning system for end-to-end noise-robust automatic speech recognition (DPSL-ASR), which is illustrated in Figure~\ref{fig1} (c).

We first examine a joint training system~\cite{liu2019jointly, pandey2021dual} in Figure~\ref{fig1}(a) by cascading the front-end SE module and back-end ASR module via multi-task learning strategy. However, the over-suppressed speech generated by speech enhancement module could significantly degrade the performance of downstream ASR task.
To alleviate this, recent study~\cite{hu2022interactive} proposed an IFF-Net to combine the enhanced Fbank feature $ X_E $ and original noisy Fbank feature $ X_N $ as a fused feature $ X_F $ to recover the over-suppressed information for ASR, as shown in Figure~\ref{fig1}(b). 
The IFF-Net has achieved significant improvement of ASR performance, but over-suppression phenomenon can still be observed in its generated fused feature $ X_F $.

To further alleviate the over-suppression problem, we first introduce clean speech feature $ X_C $ along with the fused feature $ X_F $ from IFF-Net~\cite{hu2022interactive} as dual-path inputs to the back-end ASR module, as shown in Figure~\ref{fig1}(c). 
Such parallel clean feature can provide more complementary information that has been suppressed in fused feature. 
In addition, it can also benefit the ASR training, especially at the early training stage when the poorly-trained speech enhancement module and IFF-Net cannot provide high-quality fused feature $ X_F $ for ASR. 
Then, we propose a novel style learning method to map fused feature close to the clean feature, in order to learn latent speech information from the latter, \textit{i.e.}, clean ``speech style''. 
Furthermore, we employ consistency loss to minimize the distance of ASR outputs in two paths to improve noise-robustness.
As a result, the back-end ASR module could learn more noise-robust speech representations to alleviate the over-suppression problem, and thus yield better performance. 
Since no clean data in test set is available, the clean path is discarded during inference stage.

\subsection{Style Learning}
\label{ssec:sl}
\begin{figure}[t]
  \centering
  \vspace{0.12cm}  \includegraphics[width=0.46\textwidth]{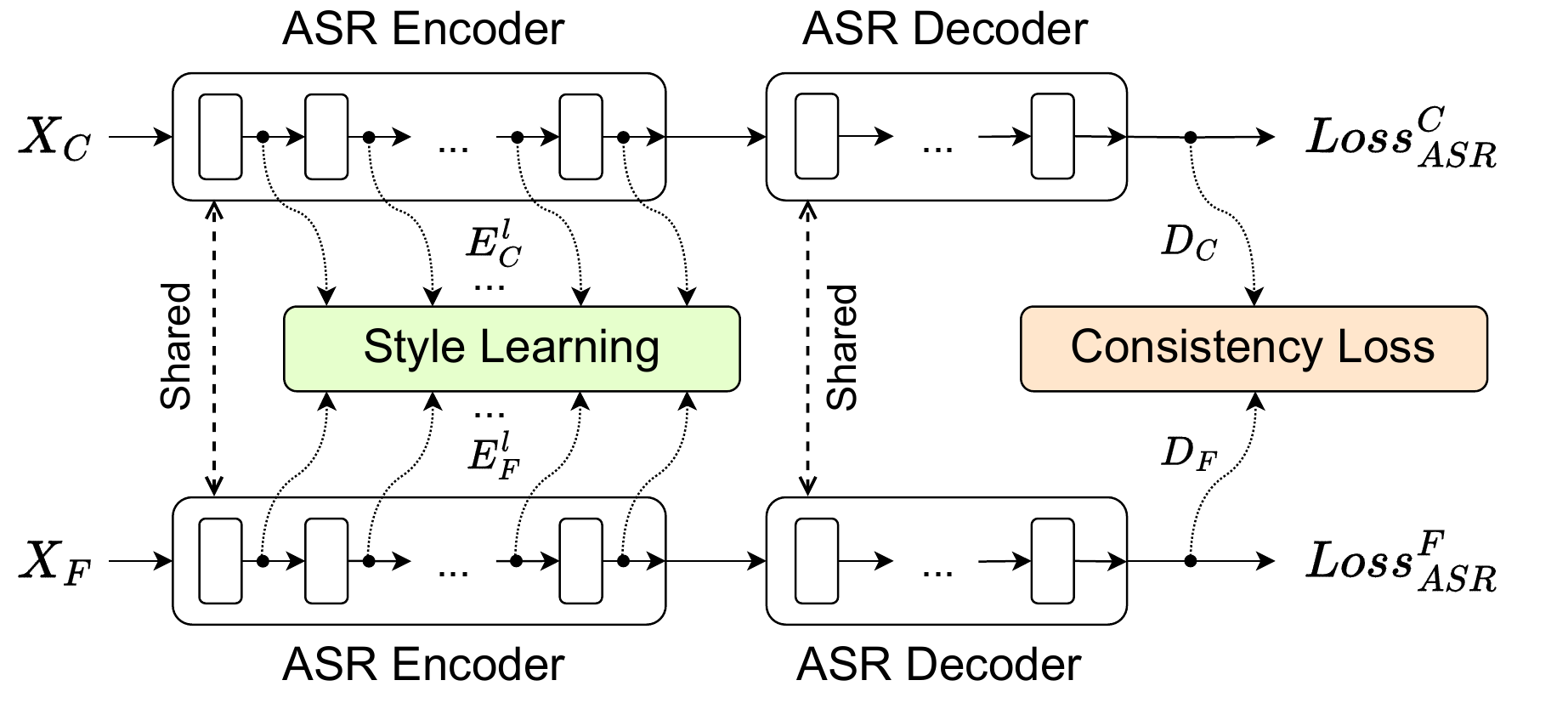}
  \vspace{-0.05cm}
  \caption{Block diagram of the back-end ASR module with style learning and consistency loss in the proposed DPSL-ASR. The dashed arrows between ASR encoders and decoders both denote sharing model parameters.}\label{fig2}
  \vspace{-0.3cm}
\end{figure}

Inspired by the image style transfer algorithm in prior work~\cite{gatys2016image}, we propose a style learning method to transfer ``speech style'' from clean encoded embeddings to fused embeddings, as shown in Figure~\ref{fig2}. 

Specifically, we first send clean feature and fused feature $ X_C, X_F \in {\mathbb{R}}^{T \times F} $ into the ASR module in two paths, where $T$ is number of time frames and $F$ is number of frequency-bins.
The ASR module in two paths share parameters with each other, which consists of $L$ Conformer layers in the encoder and $L/2$ Transformer layers in the decoder, following the prior work~\cite{gulati2020conformer}. 
Based on this, we denote the clean embedding from $l$-th encoder layer as $ {E}^{l}_{C} $ and the fused embedding from $l$-th encoder layer as $ {E}^{l}_{F}$, where $ {E}^{l}_{C}, {E}^{l}_{F} \in {\mathbb{R}}^{T \times D}$, $ l \in \{1,...,L \} $ and $D$ is embedding size. 

The ``speech style'' of clean embedding $ {E}^{l}_{C} $ and fused embedding $ {E}^{l}_{F}$ from $l$-th encoder layer are then formulated as:
\begin{equation}
\label{eq1}
\begin{split}
    {S}^{l}_{C} &= ({E}^{l}_{C})^{T} \cdot {E}^{l}_{C}, \\
    {S}^{l}_{F} &= ({E}^{l}_{F})^{T} \cdot {E}^{l}_{F},
\end{split}
\end{equation}

Equation~\ref{eq1} conducts dot product between the encoded embeddings and their own transpose, generating the style matrixes $ {S}^{l}_{C}, {S}^{l}_{F} \in {\mathbb{R}}^{D \times D} $ that indicate correlations between every two embedding channels. 
Similar to the ``image style'' learned by neural style transfer algorithm~\cite{gatys2016image}, here the calculated style matrixes contain abundant latent information inside the speech feature, which is thus defined as ``speech style''. 
Then we formulate the style loss $Loss_{SL}$ as follows:
\begin{equation}
\label{eq2}
    Loss_{SL} = \frac{1}{LD^2} \sum_{l=1}^{L} \Vert {S}^{l}_{C} - {S}^{l}_{F}\Vert_2^2,
\end{equation}
where $\Vert \cdot \Vert_2$ denotes $L_2$ norm.
Without specified, we employ the embeddings from all of $L$ encoder layers to calculate the style loss.
In this way, the ASR encoder would be optimized by style loss to learn abundant latent speech information from the clean embeddings, which could recover the suppressed information in fused embeddings and thus alleviate the existed over-suppression problem in IFF-Net.

\subsection{Consistency Loss}
\label{ssec:cl}
The consistency loss is designed to minimize the distance of outputs in two paths after ASR decoder via the symmetric Kullback-Leibler (KL) divergence~\cite{kullback1951kld, johnson2001symmetrizing}, which is formulated as follows:
\begin{equation}
\label{eq3}
    Loss_{CL} = Div_{KL}({D}_{C} \parallel {D}_{F}) + Div_{KL}({D}_{F} \parallel {D}_{C}),
\end{equation}
where $ D_C, D_F \in {\mathbb{R}}^{U \times V} $ are posterior outputs of ASR decoder (before softmax), in clean path and fused path respectively. $U$ is the number of output tokens and $V$ is the vocabulary size. 

We aim to minimize the KL-divergence-based consistency loss to push the decoded embedding in fused path closer to that in clean path. As a result, the ASR output could be more robust against the noisy and over-suppression conditions.

\subsection{Training Strategy}
\label{ssec:train_obj}
The proposed DPSL-ASR system adopts the following optimization strategy during training process:
\begin{itemize}
\item Seeking parameters of the SE module to minimize the mean square error loss of the speech enhancement task $Loss_{Enh}$;
\item Seeking parameters of the IFF-Net and ASR module to minimize the ASR loss $Loss^{C}_{ASR}$ for clean path and $Loss^{F}_{ASR}$ for fused path;
\item Seeking parameters of the ASR module to simultaneously minimize the style loss $Loss_{SL}$ for ASR encoder and the consistency loss $Loss_{CL}$ for ASR decoder. 
\end{itemize}
The DPSL-ASR system is trained in an end-to-end manner, where the overall training objective $Loss_{final}$ is formulated as:
\vspace{-0.4cm}
\begin{equation}
\label{eq4}
\begin{split}
    Loss_{final} &= (1-{\lambda}_{ASR}) \cdot {Loss}_{Enh} + {\lambda}_{ASR} \cdot {Loss}_{ASR} \\
    &+ {\lambda}_{SL} \cdot {Loss}_{SL} + {\lambda}_{CL} \cdot {Loss}_{CL} \\
    {Loss}_{ASR} &= (1-{\lambda}_{fused}) \cdot {Loss}_{ASR}^{C} + {\lambda}_{fused} \cdot {Loss}_{ASR}^{F}
    \vspace{-0.1cm}
\end{split}
\end{equation}
where $\lambda_{ASR}, \lambda_{SL}, \lambda_{CL}$ and $\lambda_{fused}$ are weighting parameters to balance different training objectives.

\vspace{-0.15cm}
\section{Experiments and Results}
\label{sec:exp_result}

\vspace{-0.1cm}
\subsection{Datasets}
\label{ssec:datasets}
\vspace{-0.15cm}
We conduct experiments on two corpora: one is Robust Automatic Transcription of Speech (RATS)~\cite{graff2014rats}, which is extremely noisy radio communication data; another is the far-field CHiME-4 dataset~\cite{vincent2016chime4} that contains normally noisy speech.

RATS dataset consists of eight parallel channels and we only use the Channel-A subset in this work, which contains 44-hour training data, 5-hour valid data and 8-hour test data. 
Although RATS dataset is chargeable by LDC, a Fbank feature version is publicly available online\footnote{https://github.com/YUCHEN005/RATS-Channel-A-Speech-Data}.

The CHiME-4 dataset consists of three subsets: clean data, real noisy data and simulated noisy data. 
The clean data comes from the WSJ0~\cite{paul1992design} training data. 
The real noisy data is recorded in four noisy environments, including bus, cafe, pedestrian area and street junction. 
The simulated noisy data is generated by mixing the clean data with background noise recorded in above four environments.
We use real and simulated noisy data of 1-channel track in this work.

\vspace{-0.25cm}
\subsection{Experimental Setup}
\label{ssec:exp_setup}
\vspace{-0.12cm}
\subsubsection{Network Configurations}
\label{sssec:network_configs}
\vspace{-0.1cm}
The proposed DPSL-ASR system consists of three modules: the SE module, the IFF-Net and the ASR module. 
The SE module is same as that in~\cite{hu2022interactive}, which utilizes 3 layer of 896-unit bi-directional long short-term memory (BLSTM)~\cite{hochreiter1997lstm}, followed by a 257-unit linear layer and a ReLU~\cite{glorot2011deep} activation function to predict masks for noisy magnitude features. 
The IFF-Net consists of 4 residual attention blocks with 64 filters, following the best configurations in prior work~\cite{hu2022interactive}. 
The ASR module contains $L=$12 Conformer~\cite{gulati2020conformer} layers in encoder, and 6 Transformer~\cite{vaswani2017attention} layers in decoder. 
The embedding dimension/feed-forward dimension/attention heads are set to 256/2048/4 for all the Conformer and Transformer layers. 
We also employ 1000 byte-pair-encoding (BPE)~\cite{kudo2018bpe} tokens as ASR output.

The network is optimized by Adam algorithm~\cite{kingma2014adam}, where the learning rate first warms up linearly to 0.002 in 25,000 steps and then decreases proportional to the inverse square root of the step number.
The batch size is set to 64. 
The training epoch is set to 50 for experiments on RATS dataset and 100 for experiments on CHiME-4 dataset.
The weighting parameters $\lambda_{ASR}, \lambda_{SL}, \lambda_{CL}$ and $\lambda_{fused}$ are set to 0.7, 0.01, 0.4 and 0.3 respectively, where we first tune $\lambda_{fused}$ to build ${Loss}_{ASR}$, and then tune $\lambda_{ASR}$ to combine ${Loss}_{Enh}$, followed by $\lambda_{SL}$ to add ${Loss}_{SL}$, and $\lambda_{CL}$ to add ${Loss}_{CL}$.
All the hyper-parameters are tuned on validation set.

\vspace{-0.25cm}
\subsubsection{Reference Baselines}
\label{sssec:baselines}
\vspace{-0.12cm}
We implement four competitive baselines to evaluate our proposed DPSL-ASR approach:
\begin{itemize}
\item \textbf{E2E-ASR}~\cite{gulati2020conformer}: a Conformer-based end-to-end ASR system.
\item \textbf{Cascaded SE-ASR}~\cite{subramanian2019speech}: a cascaded SE and ASR system that only uses final ASR loss for optimization.
\item \textbf{Joint SE-ASR}~\cite{liu2019jointly}: a cascaded SE and ASR system that employs multi-task learning strategy for joint SE-ASR training, as shown in Figure~\ref{fig1}(a).
\item \textbf{IFF-Net}~\cite{hu2022interactive}: a joint SE-ASR system with IFF-Net, as shown in Figure~\ref{fig1}(b). The IFF-Net fuses enhanced speech and original noisy speech to learn a fused representation for ASR. Therefore, the ASR module could learn complementary information to alleviate over-suppression problem.
\end{itemize}
For fair comparison, the SE modules, IFF-Nets and ASR modules in all baselines are in same structures and configurations with the proposed DPSL-ASR. 
Therefore, our DPSL-ASR contains same number of parameters as the IFF-Net baseline.

\begin{table}[t]
    \centering
    \vspace{-0.1cm}
    \caption{WER\% results of the proposed DPSL-ASR and four baselines on RATS Channel-A dataset. ``Use SL \& CL'' denotes whether use the style learning and consistency loss.}
    \vspace{-0.1cm}
    \label{table1}
    \resizebox{0.375\textwidth}{!}{
    \begin{tabular}{p{10em}|c|c}
        \toprule
        Method & Use SL \& CL & WER(\%) \\
        \midrule
        E2E-ASR~\cite{gulati2020conformer} & \textcolor{lightgray}{\xmark} & 54.3 \\
        Cascaded SE-ASR~\cite{subramanian2019speech} & \textcolor{lightgray}{\xmark} & 53.1 \\
        Joint SE-ASR~\cite{liu2019jointly} & \textcolor{lightgray}{\xmark} & 51.8 \\
        IFF-Net~\cite{hu2022interactive} & \textcolor{lightgray}{\xmark} & 46.2 \\
        \midrule
        \multirow{2}{*}{DPSL-ASR (ours)} & 
            \textcolor{lightgray}{\xmark} & 44.1 \\
          & \cmark & \textbf{41.3} \\
        \bottomrule
    \end{tabular}}
    \vspace{-0.1cm}
\end{table}

\begin{table}[t]
    \centering
    \caption{WER\% results of the proposed DPSL-ASR and four baselines on CHiME-4 1-Channel Track dataset. ``Use SL \& CL'' denotes whether use the style learning and consistency loss. ``Dev'' and ``Test'' denote WER\% results on development set and test set, respectively. ``real'' and ``simu'' denote the real noisy subset and simulated noisy subset, respectively.}
    \vspace{-0.1cm}
    \label{table2}
    \resizebox{0.48\textwidth}{!}{
    \begin{tabular}{p{10em}|c|cc|cc}
        \toprule
        \multirow{2}{*}{Method} & \multirow{2}{*}{Use SL \& CL} & \multicolumn{2}{c|}{Dev} & \multicolumn{2}{c}{Test}\\
        \cline{3-6}
         &  & real & simu & real & simu \\
        \midrule
        E2E-ASR~\cite{gulati2020conformer} & \textcolor{lightgray}{\xmark} & 8.1 & 9.6 & 14.9 & 16.1 \\
        Cascaded SE-ASR~\cite{subramanian2019speech} & \textcolor{lightgray}{\xmark} & 7.7 & 9.2 & 14.4 & 15.6 \\
        Joint SE-ASR~\cite{liu2019jointly} & \textcolor{lightgray}{\xmark} & 7.2 & 8.7 & 13.8 & 14.9 \\
        IFF-Net~\cite{hu2022interactive} & \textcolor{lightgray}{\xmark} & 6.4 & 7.9 & 12.4 & 13.4 \\
        \midrule
        \multirow{2}{*}{DPSL-ASR (ours)} & 
            \textcolor{lightgray}{\xmark} & 6.2 & 7.6 & 12.0 & 12.9 \\
          & \cmark & \textbf{5.9} & \textbf{7.2} & \textbf{11.3} & \textbf{12.2} \\
        \bottomrule
    \end{tabular}}
    \vspace{-0.3cm}
\end{table}

\vspace{-0.1cm}
\subsection{Results}
We report experiment results in terms of word error rate (WER), as our target is ASR performance while SE is just auxiliary task.

\vspace{-0.15cm}
\subsubsection{DPSL-ASR vs. Other Competitive Methods}
\label{sssec:overall_compare}
\vspace{-0.05cm}
Table~\ref{table1} summarizes the comparison between the proposed DPSL-ASR and the four baselines on RATS Channel-A dataset.
We observe that the E2E-ASR system achieves 54.3\% WER result, indicating the high difficulty of recognizing speech under extremely noisy conditions.
Compared with E2E-ASR system, the cascaded SE-ASR approach reduces 1.2\% WER absolutely by introducing speech enhancement as the front-end module to reduce noise for downstream ASR task.
Based on this, the joint SE-ASR approach obtains another 1.3\% absolute WER reduction by optimizing the SE and ASR modules simultaneously with multi-task learning strategy.
Furthermore, the IFF-Net baseline improves significantly with 5.6\% absolute WER reduction (51.8\%$\rightarrow$46.2\%), by interactively fusing the enhanced speech and original noisy speech to complement the lost information caused by over-suppression.
Finally, we observe that our proposed DPSL-ASR further alleviates this problem and obtains the best result with 10.6\% relative WER reduction over the strongest IFF-Net baseline (46.2\%$\rightarrow$41.3\%).

Table~\ref{table2} further compares the proposed DPSL-ASR with reference baselines on CHiME-4 1-Channel Track dataset. 
We observe that our DPSL-ASR approach obtains average relative WER reduction of 8.6\% over the best IFF-Net baseline, which verifies its effectiveness under normally noisy conditions.

As a result, our proposed DPSL-ASR achieves superior performance on both extremely noisy radio-channel RATS data and the normally noisy far-field CHiME-4 data.

\begin{table}[t]
    \centering
    \vspace{-0.2cm}
    \caption{WER\% results of the style learning and consistency loss with proposed DPSL-ASR on RATS Channel-A dataset. 
    ``Use SL'' denotes whether use style learning, ``Use CL'' denotes whether use consistency loss.
    The weighting parameters $\lambda_{SL}$ and $\lambda_{CL}$ are kept same as those described in Section~\ref{sssec:network_configs}.}
    \label{table3}
    \vspace{-0.1cm}
    \resizebox{0.39\textwidth}{!}{
    \begin{tabular}{p{8em}|c|c|c}
        \toprule
        Method & Use SL & Use CL & WER(\%) \\
        \midrule
        \multirow{4}{*}{DPSL-ASR} & 
            \textcolor{lightgray}{\xmark} & \textcolor{lightgray}{\xmark} & 44.1 \\
          & \cmark & \textcolor{lightgray}{\xmark} & 42.4 \\
          & \textcolor{lightgray}{\xmark} & \cmark & 43.3 \\
          & \cmark & \cmark & \textbf{41.3} \\
        \bottomrule
    \end{tabular}}
\end{table}

\vspace{-0.1cm}
\subsubsection{Effect of Style Learning and Consistency Loss}
\label{sssec:ablation}

We further report the effect of proposed style learning and consistency loss on the performance of DPSL-ASR system, as presented in Table~\ref{table3}.
Firstly we build the DPSL-ASR system with dual-path inputs, which achieves a WER of 44.1\% (vs. 46.2\% in IFF-Net baseline).
Then, we observe that using style learning method to learn clean ``speech style'' can improve the WER result by 1.7\% absolutely (44.1\%$\rightarrow$42.4\%).
Besides, applying consistency loss on decoded embeddings can also improve the performance (44.1\%$\rightarrow$43.3\%).
Furthermore, we can achieve the best result with 2.8\% absolute WER improvement (44.1\%$\rightarrow$41.3\%) by using both of them.
Therefore, we conclude that both of the proposed style learning and consistency loss contribute positively to the superior performance of DPSL-ASR, where style learning plays the dominant role.

\vspace{-0.1cm}
\subsubsection{Effect of Style Learning on Different Encoder Layers}
\label{sssec:diff_enc_layers}

\begin{table}[t]
    \centering
    \caption{WER\% results of style learning on different encoder layers, with RATS Channel-A dataset. 
    ``0'' denotes do not use style learning, ``1-3'' denotes use it on first 3 layers, ``1-12'' denotes use it on all 12 layers, etc.
    The style losses on selected encoder layers are averaged like Equation~\ref{eq2}, and the weighting parameter $\lambda_{SL}$ is kept same as before.}
    \label{table4}
    \resizebox{0.38\textwidth}{!}{
    \begin{tabular}{p{8em}|c|c}
        \toprule
        Method & Encoder Layers & WER(\%) \\
        \midrule
        \multirow{6}{*}{DPSL-ASR} & 0 & 44.1 \\
          & 1-3 & 43.9 \\
          & 1-6 & 43.4 \\
          & 10-12 & 43.3 \\
          & 7-12 & 42.7 \\
          & 1-12 & \textbf{42.4} \\
        \bottomrule
    \end{tabular}}
    \vspace{-0.2cm}
\end{table}

To understand the principles of style learning, we report the WER results of applying it on different encoder layers in Table~\ref{table4}.
We observe that the style learning on low encoder layers can improve some WER performance (44.1\%$\rightarrow$43.9\%/43.4\%), while applying it on high encoder layers achieves significantly more improvements (44.1\%$\rightarrow$43.3\%/42.7\%).
The reason could be that the low-layer encoded embeddings focus more on speech features with phonetic information, while high-layer embeddings contain richer semantic representations with linguistic information that is closer related to ASR task, so that more valuable clean ``speech style'' could be learned on high encoder layers.
The best result is achieved by performing style learning on all 12 encoder layers (44.1\%$\rightarrow$42.4\%).
Therefore, we conclude that both low-layer and high-layer style learning can improve the ASR performance, where the latter is more effective.

\vspace{-0.1cm}
\subsubsection{Visualizations of ASR Intermediate Embeddings}
\label{sssec:visualizations}

\begin{figure}[t]
  \centering
  \includegraphics[width=0.47\textwidth]{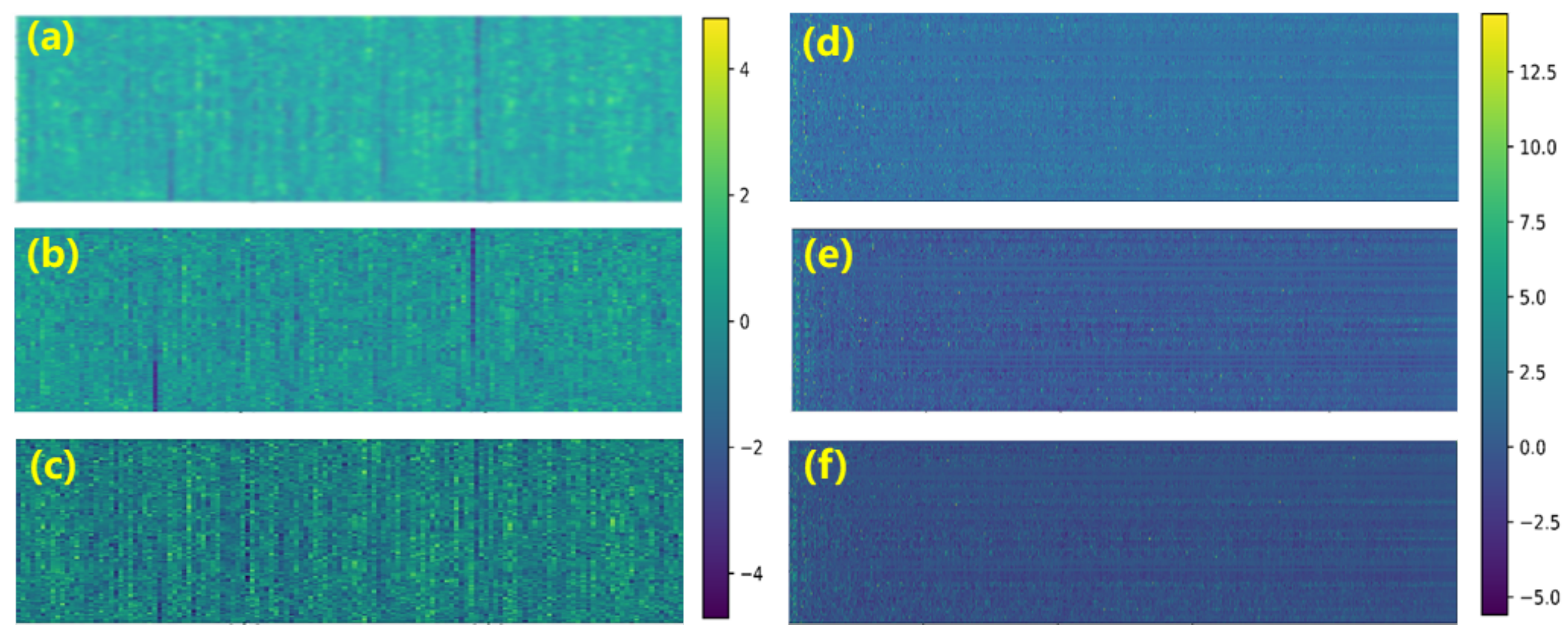}
  \vspace{-0.2cm}
  \caption{Visualizations of ASR intermediate embeddings in DPSL-ASR with a RATS test sample: encoded embeddings in (a) fused path without style learning, (b) fused path with style learning, (c) clean path without style learning; and decoded embeddings in (d) fused path without consistency loss, (e) fused path with consistency loss, (f) clean path without consistency loss.
  The x-axis denotes embedding units and y-axis denotes time frames.
  The two methods are used separately with same weighting parameters ($\lambda_{SL}, \lambda_{CL}$) as before.}\label{fig3}
  \vspace{-0.4cm}
\end{figure}

To further analyze the contribution of proposed style learning and consistency loss, we visualize the ASR intermediate embeddings in DPSL-ASR with a RATS Channel-A test sample (fe\_03\_14342-04710-A-024073-025522-A.wav), as shown in Figure~\ref{fig3}. 
We first present the encoded embeddings after final ASR encoder layer with and without style learning in (a-c). 
We observe that the encoded embedding (a) in fused path loses many latent speech information compared to (c) in clean path, where figure (a) looks blurred while (c) contains clear textures. 
After applying style learning to map encoded embeddings from fused path to clean path, the fused embedding (b) learns abundant clean ``speech style'' to alleviate over-suppression, \textit{i.e.}, it shows clearer textures than (a). 
Furthermore, we visualize the decoded embeddings after ASR decoder with and without consistency loss in (d-f). We first observe clear difference between the decoded embedding (d) in fused path and (f) in clean path, where (d) looks bright but (f) looks dark.
After applying the consistency loss, the fused embedding (e) is pushed much closer to the clean path, \textit{i.e.}, it looks darker than (d) in figure.

\section{Conclusion}
\label{sec:conclusion}
In this paper, we propose a dual-path style learning approach for end-to-end noise-robust automatic speech recognition (DPSL-ASR) to alleviate the over-suppression problem existed in prior methods.
In particular, we first introduce clean speech feature along with the fused feature from IFF-Net~\cite{hu2022interactive} as dual-path inputs to recover the over-suppressed information.
Then, we propose a style learning method to map the fused feature close to clean feature, in order to learn abundant latent speech information from the latter, \textit{i.e.}, clean ``speech style''. 
Furthermore, we also employ consistency loss to minimize the distance of final ASR outputs in two paths to improve noise-robustness.
Experimental results on RATS Channel-A and CHiME-4 1-Channel Track datasets show that the proposed DPSL-ASR approach effectively alleviates the over-suppression problem and significantly outperforms the competitive baselines.

\vfill\pagebreak

\bibliographystyle{IEEEtran}
\bibliography{mybib}

\end{document}